\documentclass[11pt]{article}
\usepackage{blois,epsfig}
\usepackage{slashed}
\usepackage{subfigure}

\def\be{\begin{equation}}
\def\ee{\end{equation}}
\def\bea{\begin{eqnarray}}
\def\eea{\end{eqnarray}}

\begin{document}
\vspace*{0.5cm}

\begin{flushright} CDF/PUB/TOP/PUBLIC/9903 \\ Version 1.2 \\
  FERMILAB-CONF-09-401-E \\
  \today \end{flushright}
\vspace*{3cm}

\title{TOP-QUARK CROSS SECTION AND PROPERTIES AT THE TEVATRON}

\author{ W. WAGNER }

\address{Bergische Universit\"at Wuppertal, Gau{\ss}stra{\ss}e 20, \\
  42119 Wuppertal, Germany}

\maketitle\abstracts{At the Tevatron, the collider experiments CDF and
  D\O \ have data sets at their disposal that compromise several hundreds
  of reconstructed top-antitop-quark pairs and allow for precision 
  measurements of the cross section and production and decay properties.
  Besides comparing the measurements to standard model predictions,
  these data sets open a window to physics beyond the standard model.
  Dedicated analyses look for new heavy gauge bosons, fourth generation
  quarks, and flavor-changing neutral currents.}

\section{Introduction}
The top quark is by far the heaviest elementary particle observed 
by particle physics experiments and features a mass of 
$m_t = 173.1\pm1.3\;\mathrm{GeV}/c^2$~\cite{:2009ec}.
The large mass the top quark gives rise to large radiative 
corrections, for example to the $W$ propagator, which causes a strong 
correlation between $m_W$, $m_t$, and the Higgs boson mass $m_H$. 
To predict $m_H$ a precise measurement of $m_t$ is crucial.
The large mass leads also to a very short lifetime
of the top quark, $\tau_t \simeq 0.5\cdot 10^{-24}\,\mathrm{s}$, such that
top hadrons are not formed. The top quark thus offers the unique 
possibility to study a quasi-free quark and as a consequence 
polarization effects are accessible in the angular distributions of 
top-quark decay products. Since $m_t$ is close
to the energy scale at which the electroweak symmetry breaks down 
(vacuum expectation value of the Higgs field $v=246\;\mathrm{GeV}$), 
it has been argued that the top quark may be part
of a special dynamics causing the break down of the symmetry~\cite{peccei}.
Finally, the top quark gives access to the highest energy scales and 
offers thereby the chance to find new, unexpected physics, 
for example heavy resonances that decay into $t\bar{t}$ pairs.

In the past years the Fermilab Tevatron, a synchrotron colliding protons and 
antiprotons at a center-of-mass energy of $\sqrt{s}=1.96\;\mathrm{TeV}$, was 
the only place to produce and observe top quarks under laboratory conditions. 
Physics data taking of Tevatron Run 2 started in 2002 and in the meanwhile the 
accelerator has delivered collisions corresponding to an integrated luminosity 
of $7.0\;\mathrm{fb^{-1}}$. The two general-purpose detectors CDF and D\O \
have recorded collisions data corresponding to $5.7\;\mathrm{fb^{-1}}$ and
$6.1\;\mathrm{fb^{-1}}$, respectively.

\section{Top-Antitop Production}
The main source of top quarks at the Tevatron is the pair production
via the strong interaction. According to the standard model (SM) top
quarks decay with a branching ratio of nearly 100\% to a bottom quark
and a $W$ boson and the $t\bar{t}$ final states can be 
classified according to the decay modes of the $W$ bosons. The most
important (or golden) channel is the so-called {\it lepton+jets} channel
where one $W$ boson decays leptonically into a charged lepton plus
a neutrino, while the second $W$ boson decays into jets.
The lepton+jets channel features a large branching ratio
of about 30\%, manageable backgrounds, and allows for the full 
reconstruction of the event kinematics. 
Other accessible channels are the {\it dilepton}
channel, where both $W$ bosons decay to leptons and the {\it all-hadronic}
channel, where both $W$ bosons decay hadronically. 

\section{Top-Antitop Cross Section}
\begin{figure}[t]
\begin{center}
\includegraphics[width=0.4\textwidth]{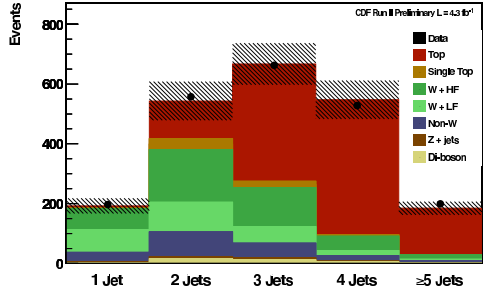}
\end{center}
\caption{\label{fig:Wjets}Jet multiplicity distribution for the 
  $W+\;$jets data set, where the $W$ boson is reconstructed in its leptonic decay
  $W^\pm \rightarrow \ell^\pm \nu_ell (\bar{\nu}_\ell)$. A cut on 
  $H_T>230\,\mathrm{GeV}$ was applied.}
\end{figure}

\begin{figure}[t]
\begin{center}
  \subfigure[]{
    \includegraphics[width=0.3\textwidth]{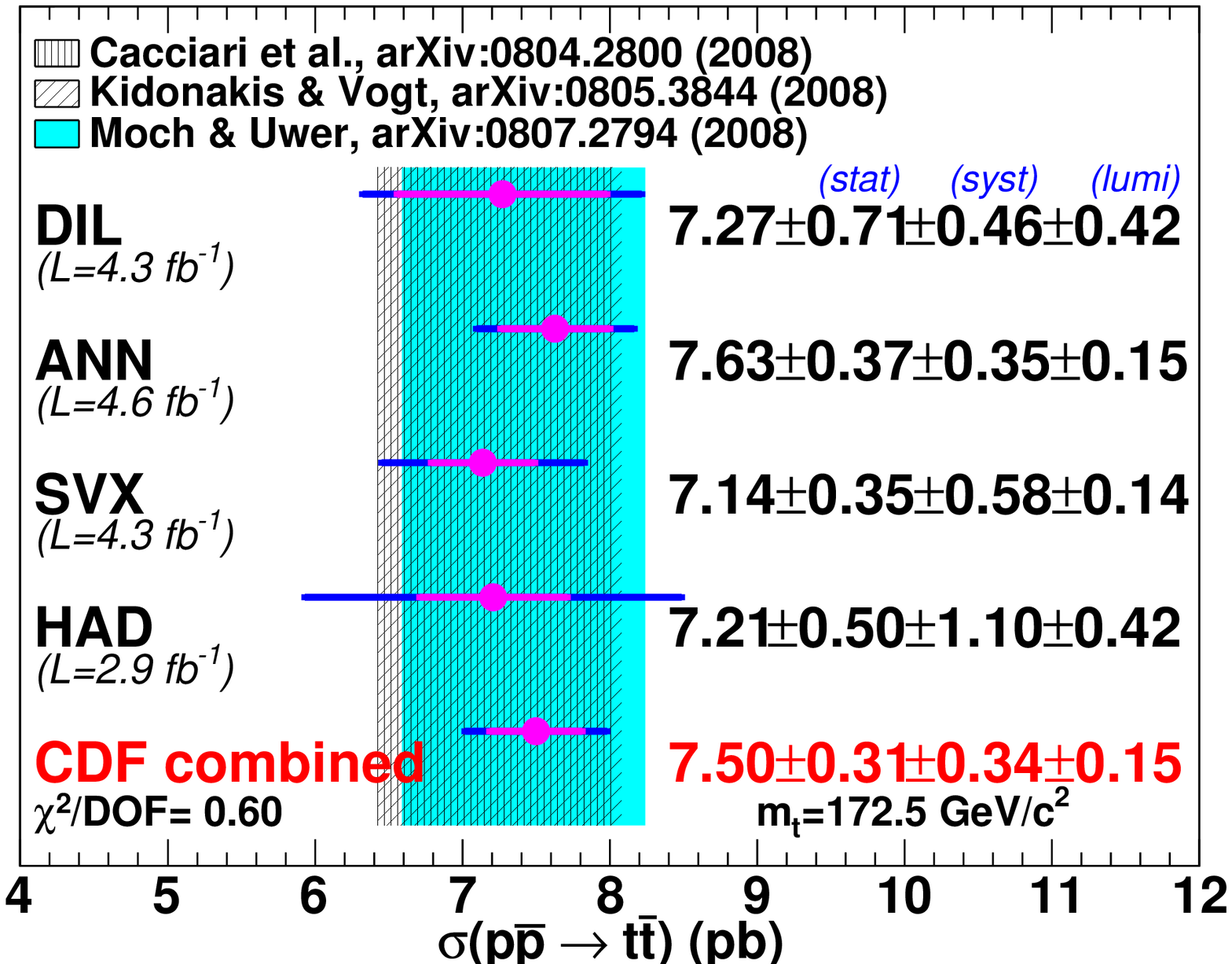}
    \label{fig:CDFcrosssections}
  }
  \subfigure[]{
    \includegraphics[width=0.25\textwidth]{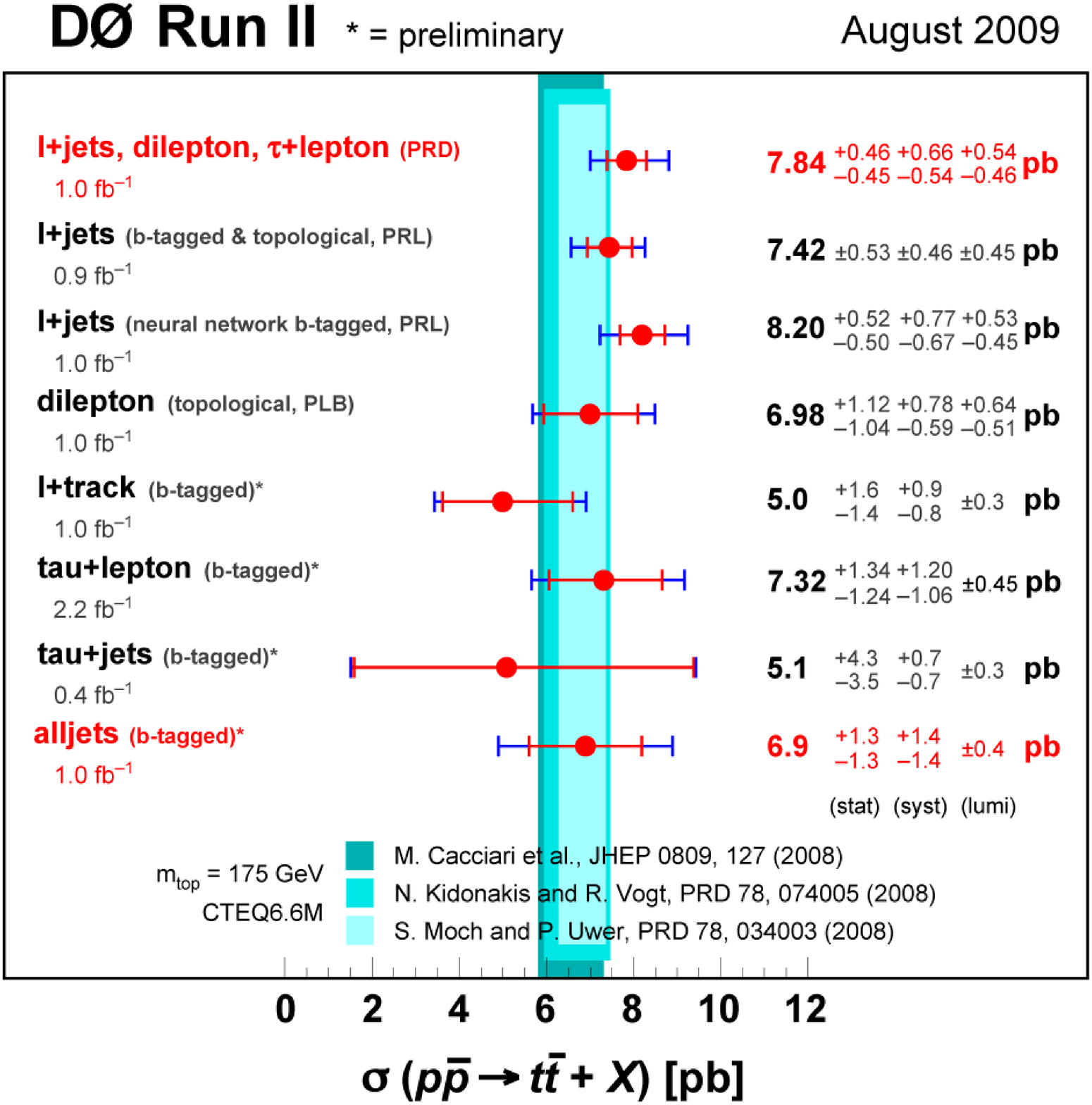}
    \label{fig:D0crosssections}
  } 
\end{center}
\caption{\label{fig:crosssections}Summary of the $t\bar{t}$ cross sections
  measured by \subref{fig:CDFcrosssections} CDF and 
  \subref{fig:D0crosssections} D\O.}
\end{figure}

The basic method to define a data sample of $t\bar{t}$ candidate events uses
the identification of $b$-quark jets by reconstructing a secondary vertex
within the jet. The corresponding CDF analysis~\cite{ttbarsecvtxCDF} is a counting 
experiment in which the background rate is estimated using a combination 
of simulated events and data driven methods. The signal region is
defined as the data set with a leptonic $W$ candidate plus $\geq 3$
jets. To further suppress background, a cut on the sum of all transverse
energies $H_T>230\,\mathrm{GeV}$ is applied. 
The jet multiplicity distribution for the $W$+jets data set observed by
this CDF analysis is shown in Figure~\ref{fig:Wjets}.
The uncertainty on the luminosity measurement is reduced by measuring
the ratio of $t\bar{t}$-to-$Z$-boson cross sections and the resulting
cross section is listed in 
Fig.~\ref{fig:crosssections}\subref{fig:CDFcrosssections}.
The single most precise measurement of the $t\bar{t}$ cross section
at CDF is based on a neural network technique applied to the 
$W+\geq 3\;$jets data set.

The cross section measurements by D\O \
are summarized in 
Fig.~\ref{fig:crosssections}\subref{fig:D0crosssections}~\cite{Abazov:2009ae}
and are interpreted by a global fit to set limits on a charged Higgs boson
that decays either in the mode $H^+\rightarrow c\bar{s}$ or
$H^+\rightarrow\tau^+\nu_\tau$. The limits are placed on the plane 
of $m_{H^+}$ vs. $\tan\beta$ and exclude the region of $\tan\beta < 2$ and
$m_{H^+}<155\,\mathrm{GeV}/c^2$ for the leptophobic mode as well as
$\tan\beta > 20$ and $m_{H^+}<155\,\mathrm{GeV}/c^2$ for the tauonic 
mode~\cite{D0H+}.

\section{Production Properties}
\paragraph{Production Mechanism}
Calculations in perturbative QCD predict that the dominating subprocess
of the production of $t\bar{t}$ pairs is $q\bar{q}$ annihilation 
(85\%), while gluon-gluon fusion contributes 15\%. At CDF, one analysis idea
to measure the fraction of $t\bar{t}$ pairs originating from a $gg$
initial state uses the proportionality of the mean number of 
low-$p_\mathrm{T}$ tracks in an event, $\bar{N}_\mathrm{trk}$, and the 
gluon content. The physical reason
for this is that $gg$ initial states produce more initial-state radiation
than $q\bar{q}$ initial states. The linear relation between 
$\bar{N}_\mathrm{trk}$ and the average number of hard initial-state 
gluons is calibrated in $W$+jets and dijet data samples.
Using simulated events, templates for the $\bar{N}_\mathrm{trk}$ 
distribution are calculated for $gg\rightarrow t\bar{t}$
and $q\bar{q}\rightarrow t\bar{t}$ events. These templates are fit
to the distribution observed in collision data, resulting in 
a measurement of 
$\sigma(gg\rightarrow t\bar{t})/\sigma(q\bar{q}\rightarrow t\bar{t})=
 0.07\pm0.14\,(\mathrm{stat.})\pm 0.07\,(\mathrm{syst.})$
\cite{:2007kq}. An alternative method exploits the spin information
in the top-decay products employing neural networks and sets an upper limit
on the $gg$ initiated fraction of $t\bar{t}$ events of
0.61 at the 95\% confidence level (C.L.)~\cite{Abulencia:2008su}.

\paragraph{Forward-Backward Asymmetry}
Due to interference effects at next-to-leading order (NLO) QCD predicts 
a forward-backward asymmetry
\[ A_\mathrm{FB} = \frac{N_t(p) - N_{\bar{t}}(p)}{N_t(p)+ N_{\bar{t}}(p)} 
   = (5.0 \pm 1.5)\%\]
at the Tevatron~\cite{afbkuhn}, where $N_t(p)$ is the number of 
top quarks moving in proton direction and $N_{\bar{t}}(p)$ is
the number of antitop quarks moving in proton direction.
Top quarks are thus more likely to be produced in proton direction,
while antitop quarks are more likely to be produced in antiproton
direction.
Using events of the lepton+jets
topology CDF and D\O \ have investigated the charge asymmetry.
In the CDF analysis the hadronic top quark is reconstructed and the
asymmetry 
\[ A_\mathrm{FB}^\mathrm{lab} = \frac{N (-Q_\ell\cdot y_\mathrm{had} > 0) - 
         N (-Q_\ell\cdot y_\mathrm{had} < 0)}
        {N (-Q_\ell\cdot y_\mathrm{had} > 0) + 
         N (-Q_\ell\cdot y_\mathrm{had} < 0)}
   = 0.193 \pm 0.065 (\mathrm{stat.}) \pm 0.024 (\mathrm{sys.})
\]        
is measured~\cite{CDFafb}. The relatively large value compared to the
SM expectation confirms earlier CDF measurements~\cite{Aaltonen:2008hc}.
The CDF measurement quoted above is corrected for background contributions, 
acceptance bias, and migration effects due to the reconstruction.

D\O \ uses $\Delta y \equiv y_t - y_{\bar{t}}$ as an observable, applies a background 
correction and obtains $A = 0.12 \pm 0.08 \pm 0.01$~\cite{afbD0}.
To compare this value with the CDF measurements or with the theory 
prediction it has to be corrected for acceptance and migration 
effects. A prescription for this is provided in reference~\cite{afbD0}.
Based on this measurement the D\O \ collaboration derives limits on a
heavy $Z^\prime$ boson that decays to $t\bar{t}$ pairs.
  
\paragraph{Top-Antitop Resonances}
The $t\bar{t}$ candidate samples offer another possibility to search for
a narrow-width resonance $X^0$ decaying into $t\bar{t}$ pairs by investigating 
the $t\bar{t}$ invariant mass. In an analysis using data corresponding to 
$3.6\;\mathrm{fb^{-1}}$ the D\O \ collaboration found no evidence for such a 
resonance and places upper limits on 
$\sigma_X\cdot\mathrm{BR}(X^0\rightarrow t\bar{t})$ range from 
$1.0\,\mathrm{pb}$ at $M_X=350\,\mathrm{GeV}/c^2$ to
$0.16\,\mathrm{pb}$ at $M_X=1000\,\mathrm{GeV}/c^2$. If interpreted in
the frame of a topcolor-assisted technicolor model these limits can be 
used to derive mass limits on a narrow lepto-phobic $Z^\prime$:
$M(Z^\prime) > 820\;\mathrm{GeV}/c^2$ at the 95\% C.L., assuming
$\Gamma(Z^\prime) = 0.012\,M(Z^\prime)$~\cite{d0Mttbar}. A similar 
analysis in the all-hadronic channel at CDF yields slightly lower mass 
limits~\cite{CDFallhadronicTT}. Another CDF analysis searches for 
a fourth generation up-type quark $t^\prime$ and sets a lower limit
of $m_{t^\prime}>311\;\mathrm{GeV}/c^2$ at the the 95\% 
C.L.~\cite{CDFtprime}.

\section{Decay Properties}
According to the SM the top quark decays with a branching
ratio of nearly 100\% to $W^+$ boson and a bottom quark via the 
weak interaction. The charged-current weak interaction has a pure
$V$-$A$ structure and thereby strongly suppresses the production of
right-handed $W^+$ bosons in top-quark decays. Only left-handed and
longitudinally polarized $W$ bosons are allowed, the production of
the later being enhanced due to the large Yukawa coupling of the 
top quark to the Higgs boson. The fraction of longitudinally polarized
$W$-bosons is predicted to be $f_0=0.70$, the left-handed fraction
$f_{-}=0.30$.

\paragraph{W Helicity}
CDF and D\O \ have measured the $W$-helicity fractions in fully
reconstructed $t\bar{t}$ lepton+jets events. 
A suitable sensitive variable to determine the $W$-helicity fractions
is the angle $\theta^*$ between the charged lepton and the negative
direction of the top quark in the $W$ rest frame.
Performing a two-dimensional and thereby model-independent fit D\O \ 
obtains $f_0 = 0.43 \pm 0.17 \pm 0.10$ and 
$f_+ = 0.12\pm 0.09 \pm 0.05$~\cite{wheld0}. Combining two similar 
analyses CDF finds $f_0 = 0.66 \pm 0.16 \pm 0.05$
and $f_+ = -0.03\pm 0.06 \pm 0.03$~\cite{Aaltonen:2008ei}.

\paragraph{Branching Ratio}
The D\O \ collaboration has recently also measured the ratio of branching
ratios $\mathcal{R} = \mathrm{BR}(t\rightarrow Wb)/\mathrm{BR}(t\rightarrow Wq)$,
an analysis which tests the hypothesis whether there is any room for an 
additional top decay channel $t\rightarrow W + q_x$ into a yet 
undiscovered quark $q_x$. In the analysis, $\mathcal{R}$ and the $t\bar{t}$
cross section are simultaneously measured~\cite{Abazov:2008yn}. The $W$+jets
data set is split in various disjoint subsets according to the number
of jets (0, 1, or $\geq 2$), the charged lepton type (electron or muon),
and most importantly the number of $b$-tagged jets. The fit results are:
$\mathcal{R} = 0.97^{+0.09}_{-0.08}$ and 
$\sigma(t\bar{t}) = 8.18^{+0.90}_{-0.84} \pm 0.50\;(\mathrm{lumi})\,\mathrm{pb}$,
where the statistical and systematic uncertainties have been combined.
The lower limit on $\mathcal{R}$ is obtained to be 
$\mathcal{R} > 0.79$ at the 95\% C.L.

\section{Conclusions}
The large data sets available in Run II of the Tevatron have propelled 
top-quark physics in a new era in which precise investigations of top-quark 
properties are possible. The top-antitop cross section  has been measured
with a relative precision of 6.4\%. Many interesting analyses have searched 
for physics beyond the SM, for example for resonances decaying into 
$t\bar{t}$ pairs. The measurements of top quark decay show impressive 
progress, most importantly the measurement of the $W$-helicity fractions in
top decay.

\section*{Acknowledgments}
The author would like to acknowledge the financial support of the 
Helmholtz-Alliance {\it Physics at the Terascale}.

\end{document}